\documentclass[epj]{svjour}
\usepackage{graphics}
\begin{document}
\title{$\eta$ Photoproduction off the nucleon revisited: Evidence for a narrow 
$N(1688)$ resonance? }

\author{V.~Kuznetsov\inst{1,2} \and M.V.~Polyakov\inst{3,4} \and T.~Boiko\inst{5}
\and J.~Jang\inst{1} \and A.~Kim\inst{1} \and W.~Kim\inst{1}\and A.~Ni\inst{1}
\and G.~Yang\inst{3}
}                     

\institute{Kyungpook National University, 702-701, Daegu, Republic of Korea,
\and Institute for Nuclear Research, 117312, Moscow, Russia,
\and Institute f\"ur Theoretische Physik II, Ruhr-Universit\"at Bochum,
D - 44780 Bochum, Germany,
\and St. Petersburg Institute for Nuclear Physics, Gatchina, 188300, St. Petersburg, Russia,
\and Belarussian State University, 220030, Minsk, Republic of Belarus.}

%
\abstract{
Revised analysis of $\Sigma$ beam asymmetry for the $\eta$ photoproduction
on the free proton reveals a structure at $W\sim 1.69$ GeV. Fit of the
experimental data based on the E429 solution of the SAID partial wave
analysis suggests a narrow ($\Gamma \leq 25$ MeV) resonance.
Possible candidates are $P_{11}, P_{13}$, or $D_{13}$ resonances. The result
is considered in conjunction with the recent evidence for a bump-like structure at
$W\sim 1.67 - 1.68$ GeV in the quasi-free $\eta$ photoproduction on the neutron.
} 
\PACS{13.60.Le\and14.20.Gk}
\maketitle

\begin{figure}
\vspace*{-0.4cm}
\centerline{\resizebox{0.4\textwidth}{!}{\includegraphics{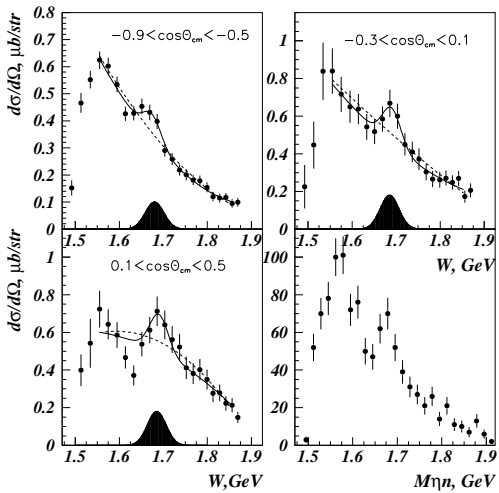}}}
\caption{Quasi-free cross sections and $\eta n$ invariant mass spectrum
for the $\gamma n \to \eta n$ reaction obtained at GRAAL (data from Ref.\protect\cite{gra1}).
Solid lines fit the data by the sum of a 3-order polynomial and a
narrow state. Dashed lines are 3-order polynomials only.
Dark areas show the simulated signals of a narrow ($\Gamma = 10$ MeV) resonance.}
\vspace*{-0.3cm}
\label{fig:etan}
\vspace{-0.3cm}
\end{figure}

The evidence for a narrow resonant structure at $W\sim 1.67 - 1.68$ in the quasi-free 
$\eta n$ photoproduction at GRAAL ~\cite{gra1,ls}, CB/TAPS@ELSA\cite{kru}, 
and LNS-Tohoku~\cite{kas} facilities is one of the most important recent findings
in the domain of the physics of nucleon resonances. 
The structure has been observed as a relatively narrow bump in the quasi-free cross section 
and in the $\eta n$ invariant mass spectrum $M(\eta,n)$ (Fig.~\ref{fig:etan}). 
Such bump is not (or poorly) seen in the   
cross-section data for $\eta$ photoproduction on the free proton~\cite{etap}.

The width of the bump in the $\gamma n \to \eta n$ 
cross section is close to that expected due to
Fermi motion of the target neutron bound in the deuteron. 
A narrow resonance which would manifest itself as a peak in the free neutron cross section, 
would appear in the quasi-free cross section as a bump of $\sim 50$~MeV(FWHM) 
width~\cite{gra1} (Fig.~\ref{fig:etan}). The $M(\eta,n)$ spectrum 
is much less affected by Fermi motion but includes larger uncertainties 
due to detector response. The observed width, of $\sim50$~MeV(FWHM), 
is  close to the instrumental resolution~\cite{gra1}.

Several attempts to explain the observed bump have been recently done.
The authors of \cite{tia,tia1,kim} suggested a narrow $P_{11}(1675)$ resonance.
Alternatively, the authors of ~\cite{skl,kl} explained the observed bump
in terms of photoexcitation and interference of the
$S_{11}(1650)$ and $P_{11}(1710)$~\cite{skl} or $S_{11}(1535)$ 
and $S_{11}(1650)$\cite{kl} resonances.

The quasi-free cross section is smeared by Fermi motion of the target neutron and is
affected by re-scattering and final-state interaction (FSI). 
Those events whose kinematics is relatively stronger distorted by Fermi motion
or those which originate from the re-scattering and FSI, are in part
eliminated in the data analysis. This procedure necessarily depends on 
experimental setup and on cuts used in data analysis. Therefore
the quasi-free cross section measured in experiment may deviate from 
the cross section calculated for the free neutron and then smeared by Fermi motion. 
$M(\eta,n)$ is almost unaffected by Fermi motion. 
Its spectrum exhibits a narrow peak at $M(\eta,n)=1.678$ GeV.  
This peak seems not to be reproduced by calculations~\cite{skl,kl}. 

Thus the bump in the $\eta$ photoproduction on the neutron 
may signal a nucleon resonance with unusual properties: 
a possibly narrow width and a much stronger photocoupling
to the neutron than to the proton. Its identification is now 
a challenge for both theory and experiment.

If photoexcitation of any resonance occurs on the neutron 
it should generally occur also on the proton, even being suppressed 
by any reason. The $\eta$ photoproduction on the proton below $W\sim 1.7$~GeV
is dominated by excitation of the $S_{11}(1535)$ resonance. 
A narrow weakly-photoexcited state with
the mass below $1.7$ GeV would appear in the cross section as a
small peak/dip structure on the slope of the dominant
$S_{11}(1535)$ resonance. This structure would be
smeared in experiment by resolution of a tagging system (for
example, the resolution of the GRAAL tagging system is 16 MeV
FWHM~\cite{pi0}), and might be hidden due to inappropriate binning.

Polarization observable - the polarized photon beam asymmetry $\Sigma$,
is much less affected by the $S_{11}(1535)$ resonance. This observable
is the measure of azimuthal anisotropy of the reaction yield when the
incoming photon is linearly polarized. 
$\Sigma$ beam asymmetry is much more sensitive to signals of
non-dominant resonances than the cross section.


For the $\eta$ photoproduction on the proton, the beam asymmetry $\Sigma$
has been twice measured at GRAAL. The first results\cite{gra2} covered
the energy range from threshold to $1.05$~GeV. 
Two statistically-independent but consistent data sets have been  reported. 
The data sets were based on two different
samples of events: i)~Events with both photons from $\eta \to 2\gamma$ decays
detected in the BGO Ball;  ii)~Events in which one of the photons, being
emitted at the angles $\theta_{lab}\leq 25^{\circ}$, was detected in the forward
shower wall, and the other was detected in the BGO ball.
The second type of events was found to be particularly efficient
at forward angles and energies above $0.9$~GeV. The results have shown
a marked peaking at forward ($\sim 40- 50^{\circ}$) angles and
$E_{\gamma} \sim 1.05$~GeV.
An extension to higher energies up to $1.5$ GeV has been reported in
Ref.\cite{gra3}. Two samples of events were merged and analyzed
together. This made it possible to significantly reduce error bars at forward angles
and to retrieve a maximum in the angular dependence at $50^{\circ}$ and
$E_{\gamma}\sim 1.05$ GeV.
A new measurement has been done by the CB/TAPS Collaboration using
a different technique of the photon-beam polarization, the coherent
bremsstrahlung from diamond radiator~\cite{bon1}. Results are in good
agreement with Refs.~\cite{gra2,gra3}.

Very recently a new data attributed to the GRAAL facility, has been
published in Ref.~\cite{ll}. This data is quite similar that presented
in Ref.~\cite{gra3} but, despite the triple increase of statistics, 
is less accurate at forward angles. The reason is that the second type of events
described above, has not been used in the data analysis\footnote{The comparison 
of $\Sigma$ beam asymmetries from Ref.~\protect\cite{gra3} and Ref.~\protect\cite{ll}
can be found in the slide 18 of Ref.~\protect\cite{gra4}.}. 
 
In Refs.~\cite{gra2,gra3,bon1} the main focus has been done on the
angular dependencies of $\Sigma$. Data points have been produced
using relatively narrow angular bins, but nearly 60~MeV wide
energy bins. Such wide bins do not allow to reveal narrow
peculiarities in the energy dependence of $\Sigma$. An ultimate
goal of this work is to produce beam asymmetries using narrow bins
in energy, in order to retrieve in detail the photon energy
dependence of $\Sigma$ for $E_{\gamma}=0.85 - 1.15$~GeV (or
$W=1.55 - 1.75$~GeV) and to search for a signal of a narrow
resonance.

In this contribution we present the revised analysis of data
collected at the GRAAL facility in 1998 - 1999. 
In general, the data analysis is the same as in Ref.~\cite{gra3}. 
Two types of events, as described above, are merged and used together 
to extract beam asymmetries.

\begin{figure}
\vspace*{0.cm}
\centerline{\resizebox{0.45\textwidth}{!}{\includegraphics{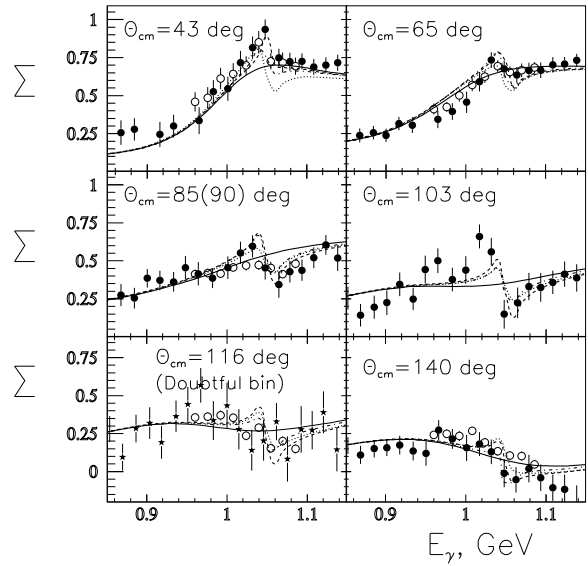}}}
\vspace*{0.cm}

\caption{Beam asymmetry $\Sigma$ for the $\eta$ photoproduction on the free proton
obtained with narrow energy bins (black circles). Open circles are data from
Ref.\protect\cite{ll}. Stars are our results at $116^{\circ}$ obtained 
using the same angular binning as in Ref.\protect\cite{ll}.
Solid lines show our calculations based on the SAID
multipoles only, dotted lines include the $P_{11}(1688)$ resonance with the
width $\Gamma=8$~MeV; dashed lines are calculations with the $P_{13}(1688)$
($\Gamma=8$~MeV), while the dash-dotted lines use the resonance $D_{13}(1688)$,
also with $\Gamma=8$~MeV.
}
\vspace*{-0.5cm}
\label{fig:as2}

\end{figure}

The results are shown in Fig.~\ref{fig:as2}. Data points are obtained using
narrow energy bins $\Delta E_{\gamma} \sim 16$ MeV. Angular bins are chosen
to be rather wide, about $(20 - 40)^{\circ}$, to gain statistics and hence
to reduce error bars.

At forward angles $(\theta_{cm}=43^{\circ})$ and $E_{\gamma}=1.04$~GeV the
data points form a sharp peak with $\Sigma$ in its maximum as large as 0.94.
The peak becomes less pronounced at $65^{\circ}$. It is
replaced by an oscillating structure at $85^{\circ}$ and $106^{\circ}$.
At more backward angles, the asymmetry above $1.05$~GeV drops down almost
to 0 (Fig.~\ref{fig:as2}) while its statistical errors grow up.

The peak at forward angles and the oscillating structure at central angles together
exhibit an interference pattern which may signal a narrow nucleon resonance. To
examine such assumption, we employ the multipoles of the recent E429 solution
of the SAID partial-wave analysis (PWA)~\cite{str1} for $\eta$ photoproduction,
adding to them a narrow Breit-Wigner resonance (as in Ref.\cite{tia1}).

The narrow $S_{11}$, $P_{11}$, $P_{13}$, and $D_{13}$ resonances are tried one by one.
Each resonance contribution is parametrized by the mass, width, photocouplings
(multiplied by the square root of the $\eta N$ branching), and the phase. These parameters
are varied to achieve the best agreement with experimental data.
The curves with the original SAID multipoles are smooth and do not exhibit any structure
(Fig.~\ref{fig:as2}). The inclusion of either $P_{11}$, or $P_{13}$, or $D_{13}$
allows to improve agreement between the data and calculations and to reproduce
the peak/dip structure. 
The mass of the included resonance is strongly constrained by
experimental data. Its value belongs to the range of $M_R=1.682 - 1.690$~GeV. The best
agreement with data corresponds to the width $\Gamma \sim 8$~MeV. However, reasonable
curves may be obtained with $\Gamma$ up to 25~MeV.

The $S_{11}$ resonance generates a dip at $43^{\circ}$ in the entire
range of variation of its photocoupling and phase. This indicates that
the observed structures, most probably, can not be attributed to a narrow
$S_{11}$ resonance.

The calculated cross section is weakly affected by the added resonances:
the $P_{11}$ generates a small peak/dip structure near $W\sim 1.69$ GeV
while the $P_{13}$ and the $D_{13}$ resonances produce almost no
effect. This explains while the possible underlying resonance is
not (or poorly) seen in the free-proton cross section data.

The mass estimate for the underlying resonance is about 5-10~MeV higher than
the value obtained in $\eta$ photoproduction on the neutron. This could be
explained by the nuclear shift of the narrow resonance mass $M$:
\begin{equation}
\Delta M = -\,\frac{<P_{F}^2>}{2M} \sim -\,5\,\,\textrm{MeV}\,,
\end{equation}
where $P_{F}$ is the Fermi momentum in the deuteron.

It is worth to noting that the authors of Ref.~\cite{ll} found
``... no evidence for a narrow $P_{11}(1670)$ state..." 
in the beam asymmetry data. In Fig.~\ref{fig:as2} our data and 
the data from Ref.~\cite{ll} are plotted together. Both data sets are consistent. 
The major difference is that we observe a dip structure at $103^{\circ}$.
The authors of Ref.~\cite{ll} show the data at $116^{\circ}$ where they do not observe 
any dip structure. However no reliable data can be produced in this ($116^{\circ}$)
angular bin. At the photon energy $1.05$ GeV recoil protons are emitted into a gap 
between the forward and the central part of the GRAAL detector where they cannot
be properly detected. A question to be addressed to the authors
of Ref.~\cite{ll}: whether do they observe a dip structure near $100^{\circ}$?
\footnote{Some consideration of this issue is given in the slide 19
of Ref.~\protect\cite{gra4}.}

In summary, we report an evidence for a narrow resonance structure
in the $\Sigma$ beam asymmetry data for the $\eta$ photoproduction
on the free proton. This structure may manifest a narrow resonance
with the mass $M\sim 1.688$~GeV and the width \mbox{
$\Gamma \leq 25$ MeV}. As candidates, narrow $S_{11}, P_{11}$,
$P_{13}$, and $D_{13}$ resonances are tried. Among them, either the $P_{11}$, or
$P_{13}$, or $D_{13}$ resonance improves the description of the data.
Most probably, the same resonance is observed in the cross section of the $\eta$
photoproduction on the neutron, where its photoexcitation is much stronger.

It is a pleasure to thank the staff of the European Synchrotron Radiation Facility
(Grenoble, France) for stable beam operation during the experimental run.
Special thanks to I. Strakovsky for many discussions, encouragement and help with
SAID data base.
We are thankful to Y.~Azimov, A.~Fix, K.~Goeke,  and L.~Tiator
for many valuable discussions.
P.~Druck is thanked for support in data processing.
This work has been supported in part by the Sofja Kowalewskaja Programme
of Alexander von Humboldt Foundation and in part by Korean Research Foundation.



\begin{thebibliography}{10}
\bibitem{gra1} V.~Kuznetsov \textit{et al.}, \textit{Phys. Lett.} B\textbf{647}, 23 (2007),
               [hep-ex/0606065]; hep-ex/0409032; hep-ex/0601002.
\bibitem{ls}   P.~Levi Sandri \textit{et al.}, Int. J. Mod. Phys. A\textbf{22}, 341 (2007).
\bibitem{kru}  H.~Schmieden, Talk at at the International Workshop on the Physics of Excited
               Baryons NSTAR2007, Bonn, Germany, September 5 - 8 2007,
	       http://nstar2007.uni-bonn.de.
\bibitem{kas}  H.~Shimizu, Talk at at the International Workshop on the Physics of Excited
               Baryons NSTAR2007, Bonn, Germany, September 5 - 8 2007,
	       http://nstar2007.uni-bonn.de. 
\bibitem{etap} F.~Renard \textit{et al.}, \textit{Phys. Lett}. B\textbf{528}, 215 (2002);
               M.~Dugger \textit{et al.}, \textit{Phys. Rev. Lett.} \textbf{89},
               222002 (2002);
               V.~Crede \textit{et al.}, \textit{Phys. Rev. Lett.} \textbf{94},
               012004 (2004).
\bibitem{tia} L.~Tiator, Int. J. Mod. Phys. A\textbf{22}, 297 (2007), [nucl-th/0610114].
\bibitem{tia1}A.~Fix, L.~Tiator, and M.~Polyakov, \textit{Eur. Phys. J.} 
               A\textbf{32} 311 (2007),[nucl-th/0702034].
\bibitem{kim} K.-~S.~Choi, S.~Nam., A.~Hosaka, and H.-~C.~Kim, \textit{Phys. Lett.}
              B\textbf{636}, 253 (2006), [hep-ph/0512136].
\bibitem{skl} V.~Shklyar, H.~Lenske and U.~Mosel, \textit{Phys. Lett}. B\textbf{650}, 172 (2007),
              [nucl-th/0611036].
\bibitem{kl}  A.~Anisovich, Talk at at the International Workshop on the Physics of Excited
               Baryons NSTAR2007, Bonn, Germany, September 5 - 8 2007,
	       http://nstar2007.uni-bonn.de.
\bibitem{pi0} General description of the GRAAL facility is available in
              V.~Bellini \textit{et al.}, \textit{Eur. Phys. J.} A\textbf{26}, 299 (2006).
\bibitem{gra2} J.~Ajaka \textit{et al.}, \textit{Phys. Rev. Lett.} \textbf{81}, 1797 (1998).
\bibitem{gra3} V.~Kuznetsov \textit{et al.}, $\pi N$ NewsLetters \textbf{16}, 160 (2002);
                Data are available in the SAID data base at \hbox{http://gwdac.phys.gwu.edu}.
\bibitem{bon1} D.~Elsner \textit{et al.}, \textit{Eur. Phys. J.} A\textbf{33}, 147 (2007), [nucl-ex/0702032].
\bibitem{ll}   O.~Bartalini \textit{et al.}, \textit{Eur. Phys. J.} A\textbf{33}, 169 (2007),
               [nucl-ex:0707.1385]. 
\bibitem{gra4}  V.~Kuznetsov \textit{et al.}, Talk at at the International Workshop on the Physics of Excited
               Baryons NSTAR2007, Bonn, Germany, September 5 - 8 2007,
	       http://nstar2007.uni-bonn.de.
\bibitem{str1} R.~A.~Arndt, W.~J.~Briscoe, I.~I.~Strakovsky, and
               R.~L.~Workman, in progress, \hbox{http://gwdac.phys.gwu.edu}.
\end{thebibliography}
\end{document}